\documentclass[12pt]{article}
\usepackage{color,multicol}
\usepackage{amsmath}
\usepackage{amssymb}
\usepackage{epsfig}
\usepackage{pifont}

\usepackage[]{hyperref}
\hypersetup{
    colorlinks = true,
    linkcolor = blue,
    anchorcolor = blue,
    citecolor = blue,
    filecolor = blue,
    urlcolor = blue
    }

\usepackage{tocloft}

\begin{document}
\title{\vspace*{-1cm}Information Geometry of Risks and Returns\footnote{Original version 9 June 2022. Journal-ref: Risk, June (2023).\vspace*{1mm}}}

\author{Andrei N. Soklakov\footnote{APAC Head of Prime and D1 Quantitative Analytics, Citibank; Andrei.Soklakov@(citi, gmail).com\vspace*{1mm}\newline
{\sl The author is very grateful to Alexander Shubert for many insightful comments. The views expressed herein should not be considered as investment advice or promotion. They represent personal research of the author and do not necessarily reflect the view of his employers, or their associates or affiliates.}
}}
\date{}
\maketitle

\begin{center}
\parbox{14.5cm}{
{\small
We reveal a geometric structure underlying both hedging and investment products. The structure follows from a simple formula expressing investment risks in terms of returns. This informs optimal product designs. Optimal pure hedging (including cost-optimal products) and hybrid hedging (where a partial hedge is built into an optimal investment product) are considered. Duality between hedging and investment is demonstrated with applications to optimal risk recycling. A geometric interpretation of rationality is presented. 
 
}
}
\end{center}

\vspace*{5mm}

\section{Introduction}
We provide a rational framework for manufacturing financial products which can be used for both hedging and investments. More precisely, we take the framework of {\it information derivatives}~\cite{Soklakov_2014_WQS}, which so far has been used for structuring investments, and extend it to the case of hedging. 

Information remains the prime underlying of optimal product design. Mathematically, it is captured by probability distributions (market-implied, investor-believed, etc.). It turns out that the same probabilistic language can be used to capture risky scenarios alongside investment views allowing for a unified approach for both hedging and investments.

The paper is organized as follows. After a short review of information derivatives, we show how to interpret all standard first order sensitivities as investment spreads (between investor-expected and market-implied rates of return). This result is interesting in its own right as it lends sharp intuition about returns~\cite{Soklakov_2018} to the understanding of risks. The same result teaches us how to compute risk with respect to a product.

We proceed by discovering a geometrical structure formed by the consensus market views, portfolio-implied views, and risk scenarios. We describe this structure showing how its shape relates to numerical risk measures. This tells us how optimal hedging products should be designed. We consider cost-optimal hedging, pure hedging (which express as little view as possible) and learn how to incorporate partial hedging within optimal investment products. We also reveal a certain duality between hedging and investments. We show, for instance, how an optimal hedge can be implemented by offering an optimal investment to a suitably chosen client (optimal risk recycling).

As a bonus, we obtain a geometric theory of rational behaviour. In this theory risk averse actors are trying to be as close as possible to the market. The actors are constrained by their views but can use different measures to quantify their divergence from the market. The net behaviour is equivalent to the actors maximizing their expected utilities.

Detailed calculations and interdisciplinary notes are deferred into a set of appendices. These can be skipped in the first reading and consulted depending on interest.

\section{Information derivatives}\label{Sec:InformationDerivatives}
\subsection{Multiple rationality}
Behaviour of a human person is very complex and, as far as we know, it does not follow any single-goal optimization. At first sight, this seems to confirm the well-known research narrative that individual people are fundamentally irrational.

On closer inspection, however, many apparently irrational behaviours do appear to have reasons. It is perhaps more accurate to think of individuals as {\it multi-rational} rather than simply irrational (see Appendix~\ref{Sec:MultipleRationality}).

Multiple rationality suggests that even a single person is driven by many goals. Indeed, a person consumes many different products judging them with respect to the specific needs the products are designed to serve.

Multiple rationality is interesting to us because it suggests that the relatively old mathematics of rationality, such as the expected utility formalism, might be applicable to the design of some products even if it fails to capture the full complexity of human behaviour. 

\subsection{Science of product design}
Mathematically, {\it financial products} are defined by their payoff functions which state how benefits (normally cashflows) depend on the underlying variables. 

Without loss of generality, we can assume that all individual payoff functions are non-negative, i.e. that individual financial products are {\it assets}. Indeed, one can always take apart any venture into assets and liabilities and consider the liabilities as a result of shorting assets. 

Furthermore, variable amounts of capital can be invested in the same product without changing its nature. The payoff function is therefore defined up to an arbitrary multiplier (commonly called a notional).

Inspired by multiple rationality, one can attempt to build financial products as solutions to a great variety of possible optimizations. However, only a small fraction of theoretical constructions are relevant in practice. This calls for a scientific theory of product design, i.e.~a theory which seeks consistency with observed facts.\\ 

An example of that is the framework of {\it information derivatives}. Mathematically, it is built around the basic laws of information processing and rational optimizations. Scientifically, the framework is testing itself against econometrics and neuroscience~\cite{Soklakov_2018, Soklakov_2014EqPuzzle}. 

The simplest example of an information derivative is the {\it likelihood product} which naturally arises in the context of Bayesian learning~\cite{Soklakov_2011}. More formally, let $m(x)$ be the {\it market-implied} distribution for some underlying variable $x$, and let $b(x)$ denote the {\it investor-believed} distribution for the same variable. One can think of $m(x)$ as the {\it prior information} about $x$ that is already priced by the market and interpret $b(x)$ as the {\it posterior information} which describes the investor's belief post research. The payoff $f(x)$ of the {\it likelihood product} is defined as the likelihood function which connects the prior and the posterior distributions according to the familiar product rule
\begin{equation}\label{Eq:b=fm}
b(x)=f(x)\,m(x)\,.
\end{equation}
The simplest interpretation of this equation as Bayes' rule justifies the terminology. At the same time Eq.~(\ref{Eq:b=fm}) admits many practical inference frameworks (see Appendix~\ref{Sec:InferenceProduct}).

It turns out that the concept of a likelihood product (or a {\it likelihood investor}\footnote{We use the terms {\it investor} or {\it hedger} as metonyms of the relevant {\it product} (or {\it strategy}).}) is very useful for understanding a large class of investment strategies. This observation is captured by the {\it investor equivalence principle}~\cite{Soklakov_2013b}. 

The principle contains a trivial mathematical statement and a deeper scientific part. Mathematically, the principle takes any product (bought by any investor) and imagines an ({\it economically equivalent}\/) likelihood investor who happens to buy the exact same product. This can always be done. Indeed, consider an arbitrary product $F(x)$, and set
\begin{equation}\label{Eq:InvestorEquivalence}
\beta_F(x)= F(x)\,m(x)\,.
\end{equation}
Since the payoff $F(x)$ is defined up to a notional multiplier (as is the likelihood function), one can make sure that $\beta_F(x)$ is a probability distribution and interpret $F(x)$ as the likelihood product corresponding to the view $\beta_F$. 

The investor equivalence principle encourages us to think in terms of likelihood products. Therein lies its scientific content (beyond pure mathematics), which helps us to focus on realistic strategies~\cite{Soklakov_2013a}. To demystify how this might happen, note that (in most practical circumstances) likelihood products maximize expected returns~\cite{Soklakov_2011}.\footnote{This interpretation breaks down for large transactions with significant impact on the market.} Thinking in terms of returns captures important aspects of human decision making~\cite{Soklakov_2018}. Returns influence implementation, maintenance, and termination of practical strategies.
 
All results of this paper are obtained by frequent use of the investor equivalence principle. As a practical point, before proceeding, we want the reader to be comfortable using Eq.~(\ref{Eq:InvestorEquivalence}) to map payoff functions into implied views and vice versa ($F\leftrightarrow\beta_F$).

The likelihood product is a convenient steppingstone from which we can reach a large class of rational strategies. Using the likelihood function $f$ as a benchmark, the equation for a more general rational product $F$ reads~\cite{Soklakov_2013b} 
\begin{equation}\label{Eq:PayoffElasticity}
\frac{d\,\ln F}{d\,\ln f}=\frac{1}{R}\,,
\end{equation}
where $R$ is the investor's relative risk aversion. Mathematically, the payoff elasticity equation~(\ref{Eq:PayoffElasticity}) is just the necessary (Euler-Lagrange) condition which $F$ must satisfy to maximize the investor-expected utility $\int b(x) U(F(x))\,dx$. In this notation $R=-F U''_{FF}/U'_F$ (the Arrow-Pratt relative risk aversion). 

It is important to remember that Eq.~(\ref{Eq:PayoffElasticity}) describes just a single rational strategy and not the overall behaviour of an entire human person (let alone an economy).\footnote{Every person is playing host to many strategies (see Appendix~\ref{Sec:MultipleRationality}). The individual strategies can have contradicting goals and compete for limited resources (even within the scope of a single person). Some strategies manage to spread between people. In extreme cases, popular strategies can even organize people, compelling them to create specialized infrastructure such as equity exchanges.}

\section{Risks as returns}\label{Sec:RisksAsReturns}
Consider a portfolio of derivative assets with a common underlying variable $x$ and combined payoff function $\Pi(x)\geq 0$. The price of the portfolio can be written as
\begin{equation}\label{Eq:Price}
{\rm Price}[\Pi]=\int\Pi(x)\,m_\sigma(x)\,dx\,,
\end{equation}
where $m_\sigma(x)$ is the market-implied distribution for $x$ and where $\sigma$ is some parameter. The sensitivity of the price with respect to the parameter is
\begin{equation}\label{Eq:Sensitivity}
\partial_\sigma{\rm Price}[\Pi]=\int\Pi(x)\,\frac{\partial m_\sigma(x)}{\partial\sigma}\,dx\,.
\end{equation}

From the point of view of an investor a small variation $\delta$ of the parameter $\sigma$ is captured by the view
\begin{equation}
b(x)=m_{\sigma+\delta}(x)\,,
\end{equation}
and the corresponding likelihood product~(\ref{Eq:b=fm})
\begin{equation}
f(x)=\frac{m_{\sigma+\delta}(x)}{m_\sigma(x)}\,.
\end{equation}
In the simplest case of constant $R$, equation~(\ref{Eq:PayoffElasticity}) gives us the optimal investment product (unit price)
\begin{equation}\label{Eq:FRdelta}
F_R^\delta = \frac{f^{1/R}}{{\rm Price}[f^{1/R}]} = \frac{(m_{\sigma+\delta}/m_{\sigma})^{1/R}}{{\rm Price}\Big[(m_{\sigma+\delta}/m_{\sigma})^{1/R}\Big]}\,.
\end{equation}
In the risk-neutral limit ($\delta,R\to 0$) this becomes the exponential score product
\begin{equation}\label{Eq:ScoreProduct}
F_0\,\stackrel{\rm def}{=}\,\lim_{\epsilon\to 0}F_\epsilon^\epsilon\, =\, \frac{e^{{\rm Score}}}{{\rm Price}[e^{\rm {\rm Score}}]}\,,\ \ \ \ {\rm Score}(x)=\frac{\partial\ln m_\sigma(x)}{\partial\sigma}\,.
\end{equation}
Equation~(\ref{Eq:InvestorEquivalence}) applied to the portfolio $\Pi$ defines
\begin{equation}\label{Eq:BetaPi}
\beta_\Pi(x)=\frac{\Pi(x)}{{\rm Price}[\Pi]}\,m_\sigma(x)\,.
\end{equation}
The sensitivity per unit price (lets call it {\it specific sensitivity}) becomes (see Appendix~\ref{Sec:SpecificRisk})
\begin{equation}\label{Eq:SpecificSensitivity_spread}
\frac{\partial_\sigma{\rm Price}[\Pi]}{{\rm Price}[\Pi]} = E_{\beta_{\Pi}}[\ln F_0] - E_{m_\sigma}[\ln F_0]\,.
\end{equation}
This equation understands the usual price sensitivity as a spread in expected returns. One of the expectations, $E_{\beta_{\Pi}}$, is what a likelihood (growth-optimizing) investor into $\Pi$ would compute. The other, $E_{m_\sigma}$, is just the market-implied expectation. The product $F_0$ is independent from the portfolio $\Pi$. The role of $F_0$ is to capture the market scenario the sensitivity to which we perceive as risk. 

The ability of investment products to capture market scenarios suggests the following general definition of risk. The {\it specific risk} (i.e. risk per unit price) of $\Pi$ with respect to product $S$ is defined as the spread in expected returns
\begin{equation}\label{Eq:SpecificRisk}
\frac{{\rm Risk}_S[\Pi]}{{\rm Price}[\Pi]} \stackrel{\rm def}{=} E_{\beta_{\Pi}}[\ln S] - E_{m}[\ln S]\,.
\end{equation}

In the special case of the exponential score product~(\ref{Eq:ScoreProduct}), i.e. when $S=F_0$, this general definition reduces to the ordinary sensitivities~(\ref{Eq:SpecificSensitivity_spread}).
Note also that the general definition need not mention any explicit parameters (such as $\sigma$ used above). Indeed, the scenario underlying the risk is captured in its entirety by the structure of the product $S$.

The above understanding of risks in terms of returns is already quite promising. The human brain is indeed very sensitive to financial returns. This can be seen by examining ordinary marketing materials for retail customers; these often specify mortgage or savings rates within a small fraction of a percent over an annual horizon. The ability to feel risks within that kind of accuracy is interesting in its own right. For a more detailed discussion on harnessing returns-based intuition and some relevant neuroscience see Ref.~\cite{Soklakov_2018}.

\section{Information geometry of risk}

Accurate intuition is key to practical decision making. Mapping abstract sensitivities into financial returns (see above) is one possibility to enhance intuition. Using basic geometry is another. With a bit of calculus Eq.~(\ref{Eq:SpecificRisk}) can be rearranged as follows (Appendix~\ref{Sec:SpecificRisk_RelativeEntropy})
\begin{equation}\label{Eq:RiskPythagoras}
\frac{{\rm Risk}_S[\Pi]}{{\rm Price}[\Pi]}= D(\beta_\Pi\,||\,m)+D(m\,||\,\beta_S)-D(\beta_\Pi\,||\,\beta_S)\,,
\end{equation}
where $\beta_S=Sm$, and $D(p\,||\,q)$ is the Kullback-Leibler divergence (relative entropy)
\begin{equation}
D(p\,||\,q)\stackrel{\rm def}{=}\int p(x)\ln\frac{p(x)}{q(x)}\,dx\,.
\end{equation}
Equation~(\ref{Eq:RiskPythagoras}) describes risk as a property of the triangle formed by three distributions: $\beta_\Pi$, $m$ and $\beta_S$ (which define respectively: the view expressed by our portfolio, the market-implied distribution and the risk scenario). 

The quantity on the right-hand side of Eq.~(\ref{Eq:RiskPythagoras}) is closely connected to the Pythagorean theorem. This connection is in fact a cornerstone within the field of {\it information geometry} which studies families of probability distributions~\cite{Chentsov_1968, Amari_2016}. 

\begin{center}
\includegraphics[width=\textwidth]{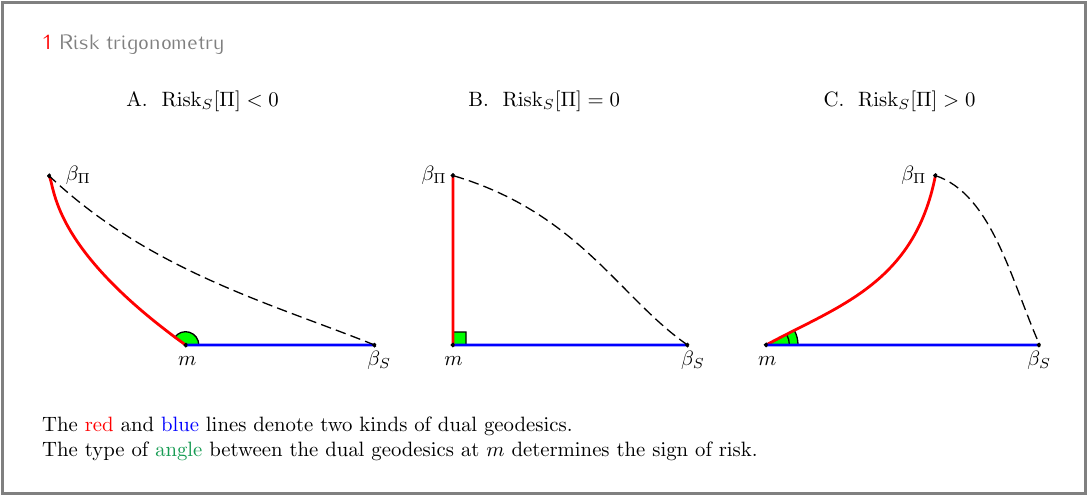}
\end{center}

The individual divergences on the rhs of Eq.~(\ref{Eq:RiskPythagoras}) play a role akin to squared distances. This is most easily seen in the limit of a very small triangle (when $\beta_\Pi$, $m$ and $\beta_S$ differ from each other by infinitesimal variations of parameters). For large triangles the asymmetry of divergences becomes significant. In general, we introduce two kinds of geodesics forming an angle at point $m$ (see Fig.~1). For a detailed review on the geometry of such triangles see Ref.~\cite{Nielsen_2021}.

When the angle at $m$ is a right angle the sum of the divergences $D(\beta_\Pi\,||\,m)+D(m\,||\,\beta_S)$ exactly compensate the divergence on the hypotenuse $D(\beta_\Pi\,||\,\beta_S)$. Given by Eq.~(\ref{Eq:RiskPythagoras}), this is the case of zero risk (Fig.~1.B).  

Positive risk implies that the angle at $m$ is acute (Fig. 1.C). This is a mathematically precise way of saying that our portfolio and the risk scenario are on the \lq\lq same side'' relative to the market. Negative risk implies an obtuse angle indicating the portfolio and the scenario are on \lq\lq opposite sides'' relative to the market (Fig.~1.A).

Let us take a closer look at the geodesics which form the above (risk-discriminant) angle at $m$. To this end let us consider the following couple of one-parameter families 
\begin{eqnarray}
p_{\rm mix}(x,t) &=& (1-t)m(x) +t\beta_\Pi(x) \,, \label{Eq:m-geodesic} \\
p_{\rm exp}(x,t) &=& \exp\Big((1-t)\ln m(x)+t\ln \beta_S(x)-\psi(t)\Big)\,, \label{Eq:e-geodesic}
\end{eqnarray} 
where the mixture family $p_{\rm mix}(x,t)$ interpolates between $m$ and $\beta_\Pi$ while the exponential family $p_{\rm exp}(x,t)$ connects $m$ with $\beta_S$. In both cases $t\in[0,1]$ is a free parameter, and $\psi(t)$ ensures the normalization of $p_{\rm exp}(x,t)$.

The term {\it geodesic} in information geometry denotes an affine generalization of a straight line. One can see how the above families form straight lines by introducing dual coordinate systems and computing two kinds of tangent vector~\cite{Chentsov_1968, Amari_2016}:
\begin{eqnarray}
\langle m, \beta_\Pi| &\stackrel{\rm def}{=}& \frac{d}{dt}\big( p_{\rm mix}\big) = \beta_\Pi - m \,, \label{Eq:bra}\\
|m, \beta_S\rangle &\stackrel{\rm def}{=}& \frac{d}{dt}\big(\ln p_{\rm exp} +\psi\big) = \ln \beta_S - \ln m\,. \label{Eq:ket}
\end{eqnarray}

\begin{center}
\includegraphics[width=\textwidth]{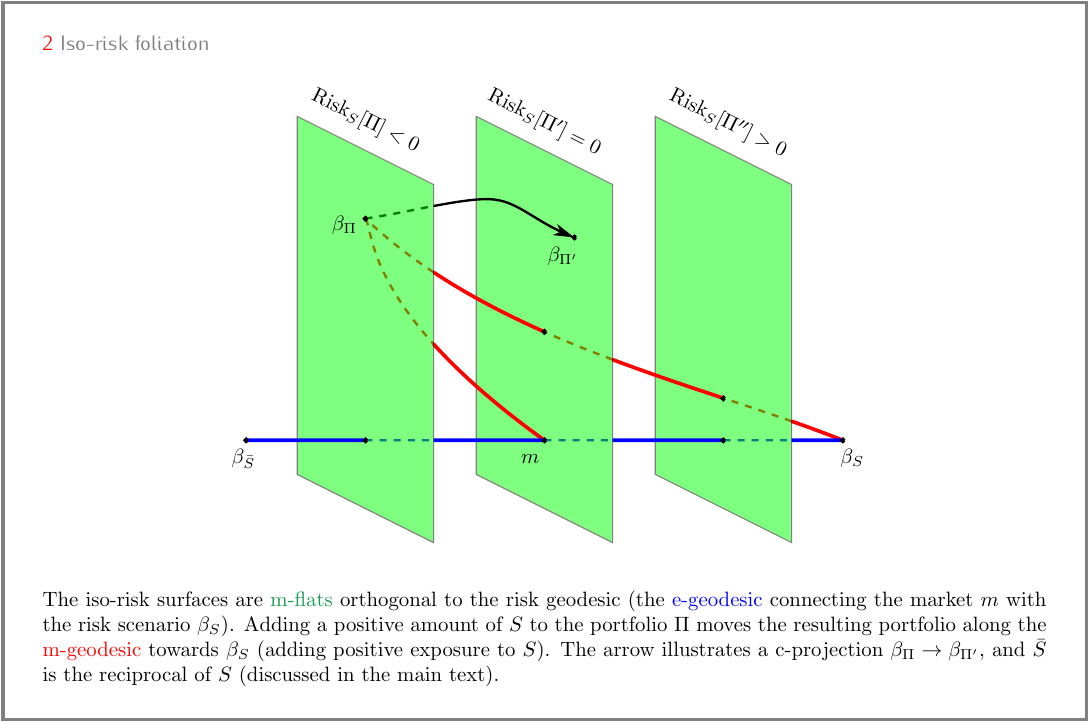}
\end{center}

Note that these tangent vectors are $t$-independent. The mixture family $p_{\rm mix}(x,t)$ becomes a straight line in the coordinates which represent a distribution as a vector of probabilities. The dual coordinates, in which the exponential family $p_{\rm exp}(x,t)$ traces out a straight line, are logarithmic in probabilities. These are classic examples of dual geodesics in information geometry. 

The mixture geodesics ({\it m-geodesics} for short) and the exponential geodesics ({\it e-geodesics}) illustrated above have clear financial interpretations. Moving along $p_{\rm mix}$ from $\beta_\Pi$ in the direction of the market $m$ is equivalent to converting a portion of the portfolio into risk-free cash (at market prices with no transaction costs). Moving along $p_{\rm exp}$ from $\beta_S$ to $m$ attenuates the risk scenario in the same way as a scaled-up risk aversion attenuates an investor's view (see Appendix~\ref{Sec:GeodesicsFinMeaning} for details).

The scalar product between tangent vectors leads naturally to the notion of an angle between geodesics~\cite{Nielsen_2021}. In the above case of dual geodesics we have
\begin{equation}
\langle m,\beta_\Pi|m,\beta_S\rangle = \int\Big(\beta_\Pi(x)-m(x)\Big)\Big(\ln\beta_S(x)-\ln m(x)\Big)\,dx\,.
\end{equation}
Comparison with Eq.~(\ref{Eq:SpecificRisk}) shows that this scalar product is exactly the specific risk
\begin{equation}\label{Eq:RiskAsScalarProduct}
\langle m,\beta_\Pi|m,\beta_S\rangle=  \frac{{\rm Risk}_S[\Pi]}{{\rm Price}[\Pi]}\,.
\end{equation}
The geometric intuition embedded in this equation allows us to see the structure formed by the surfaces of constant risk. Imagine varying $\beta_\Pi$ while keeping the scalar product~(\ref{Eq:RiskAsScalarProduct}) constant. One can correctly guess that the iso-risk surfaces are intersecting the e-geodesic $|m,\beta_S\rangle$ at right angles as shown in Fig.~2 (see Appendix~\ref{Sec:IsoRiskFoliation} for details). 

\section{Hedging with information derivatives}
The above geometric intuition tells us how to approach hedging. A portfolio $\Pi$ of assets with ${\rm Risk}_S[\Pi]\neq 0$ corresponds to the growth-optimal view $\beta_\Pi$ which lies outside the zero-risk surface (see e.g.~Fig.~2). The purpose of hedging is to eliminate risk, i.e. to move from $\beta_\Pi$ to a new location on the zero-risk surface. This can be done in a variety of ways.

Let us take a portfolio $\Pi$ of assets with ${\rm Risk}_S[\Pi]<0$. Adding to the portfolio some positive amount of $S$ would move us along the m-geodesic from $\beta_\Pi$ in the direction of $\beta_S$ (see Fig.~2). This geodesic intersects the zero-risk surface showing that we can use $S$ as a hedge. 

In the above example, the choice of $S$ as a hedge is natural but not at all unique. We could pick a different point $\beta_t$ on the (extrapolated) e-geodesic $\beta_t = p_{\rm exp}(x,t)$. The exposure of the resulting hedge to $S$ is monotonic along the e-geodesic. Indeed, with a slight abuse of notation (see Appendix~\ref{Sec:ExposureRiskGeodesic})
\begin{equation} \label{Eq:RiskAsIntegratedVar}
{\rm Risk}_S[\beta_t]=\int_0^t{\rm Var}_{\beta_\tau}[\ln S]\,d\tau\,,
\end{equation}
where ${\rm Var}_{\beta_\tau}[\ln S]$ is the variance of the log-return on $S$ (according to $\beta_\tau$). 

Analytical simplicity of the e-geodesic~(\ref{Eq:e-geodesic}) together with the monotonicity of risk~(\ref{Eq:RiskAsIntegratedVar}) make looking for a hedge with a desired amount of risk an easy linear search along the geodesic. The positive values of $t$ provide a line of products that can be bought as a hedge for negative risk (as per the above example) while the negative values of $t$ perform the exact same job for hedging positive exposure to $S$.

We would like to do more, however. We want to optimize hedging in a variety of situations. We may also want to sell our risks (as opposed to quell them by buying insurance assets). The following notes touch upon such applications.

\subsection{Hedging as optimal transfer}

Among all possible moves $\beta_\Pi \to \beta_{\Pi'}$ to the zero-risk surface the ones with the lowest cost are of prime interest. Let us call such moves {\it c-projections} of $\beta_\Pi$ onto the zero-risk manifold (Fig.~2). The problem of moving a distribution in the presence of costs inspired the entire field of {\it optimal transport}\/~\cite{Villani_2009}. Let us understand what exactly needs transporting in our case and at what cost. 

Take a portfolio of unit worth (${\rm Price}[\Pi]=1$). From Eq.~(\ref{Eq:BetaPi}) we can interpret $\beta_\Pi(x)$ as a distribution of funds across the outcomes $\{x\}$. The move $\beta_\Pi\to\beta_{\Pi'}$ transports the funds between the outcomes. In ideal settings one can think of these outcomes as the Arrow-Debreu securities. 

Selling one security and buying another attracts instantaneous \lq\lq local'' costs (independent of how far apart the two securities appear to be). The total cost of hedging the unit portfolio (${\rm Price}[\Pi']={\rm Price}[\Pi]=1$) can be modelled as a sum of the relevant local contributions
\begin{equation}\label{Eq:Cost}
{\rm Cost}(\beta_\Pi\to\beta_{\Pi'})=\int C\big( x,\beta_{\Pi'}(x)-\beta_\Pi(x) \big)\, dx\,.
\end{equation}  
The first argument of the integrand reflects the costs' local properties (e.g. the dependence on liquidity of the tradable events $\{x\}$). The second argument captures the dependence of costs on the amount of trading. The costs start at zero $C(x,0)=0$ when no trading is necessary and increase $C(x,y)>0$ for both buying $y>0$ and selling $y<0$. We assume that higher trading volumes attract increasingly higher costs, i.e. that $C(x,y)$ is smooth and convex with respect to its second argument.

Solving for the c-projection (see Appendix~\ref{Sec:C-projection}) gives us the general structure of the hedge 
\begin{equation}\label{Eq:c-proj_Mx}
\Pi'(x)-\Pi(x) = M_x\big(S(x)\big)\,,
\end{equation}
where $M_x()$ is a monotonic map. In summary, a cost-optimal hedge for the risk with respect to product $S$ is similar to $S$ up to a local monotonic map.

\subsection{Pure hedging products}
\label{Sec:HedgingProducts}
Hedging products are constrained by the combination of risks they are required to have. Any product with non-zero risk necessarily expresses some view on the market. We call a hedging product {\it pure} if it expresses as little a view as possible -- just enough to implement the required combination of risks.

For example, consider a unit worth of a hedging product $H(x)$ with risk
\begin{equation}\label{Eq:RiskOfH}
{\rm Risk}_{S}[H]=r\,.
\end{equation}
Using this as a constraint we want the view
\begin{equation}
\beta_H(x) = H(x)\, m(x)
\end{equation}
to be as close as possible to the market-implied $m$. In the above geometric picture, the obvious solution is to locate $\beta_H$ (and with it the required hedge $H$) somewhere on the e-geodesic $|m,\beta_S\rangle$. As discussed above, this is achieved by setting $\beta_H=\beta_t$ and finding the value of $t^*$ such that ${\rm Risk}_{S}[\beta_{t^*}]=r$. Found in this way, $\beta_H$ is called the e-projection $\beta_{t^*}$ of $m$ onto the iso-risk manifold~(\ref{Eq:RiskOfH}).

Iso-risk manifolds are m-flat (i.e. they can be imagined as sheets spanned by m-geodesics). More precisely, any m-geodesic connecting two points of an iso-risk manifold lies entirely within the manifold. This is a geometric manifestation of the fact that a portfolio of products with the same specific risk has the same specific risk as its constituent products. 

The number of constraints of the form~(\ref{Eq:RiskOfH}) does not change this geometric picture. Any collection of such constraints defines an m-flat (an intersection of m-flats is m-flat). This is interesting because the e-projection onto any m-flat is always unique and can be found numerically in a variety of ways including the purely geometric algorithm by Csisz\'ar~\cite{Csiszar_1975}. 

Returning to the single-risk case for simplicity, the reader can check that finding the pure hedge $H$ via the e-projection $\beta_{t^*}$ of $m$ onto the iso-risk manifold~(\ref{Eq:RiskOfH}) is equivalent to the minimization
\begin{equation}\label{Eq:KL-projection}
\beta_H = \arg \min_\beta D(\beta\,||\,m)\,,\ \ \ {\rm Risk}_{S}[\beta]=r\,.
\end{equation}
This appears to be quite specific, so let us see what happens if we replace the relative entropy in~(\ref{Eq:KL-projection}) with a more general notion of the $\phi$-divergence~\cite{Csiszar_1963, Morimoto_1963, AliSilvey_1966}
\begin{equation}\label{Eq:PhiDivergence}
D_\phi(\beta\,||\,m)\stackrel{\rm def}{=}\int m(x)\,\phi\Big(\frac{\beta(x)}{m(x)}\Big)\,dx\,,
\end{equation}
where $\phi$ is a strictly convex function such that $\phi(1)=0$. The risk constraint~(\ref{Eq:RiskOfH}) ensures that the optimal $\beta_H$ is still on the relevant iso-risk flat, but it may be displaced from the point $\beta_{t^*}$ where the e-geodesic $|m,\beta_S\rangle$ intersects the flat.

Upgrading $D$ with $D_\phi$ in~(\ref{Eq:KL-projection}) gives us the general structure of the optimal pure hedge (see Appendix~\ref{Sec:Pure_hedging_products_gen_structure})
\begin{equation}\label{Eq:OptimalPureHedge_GeneralForm}
H(x)=M_\phi\big(S(x)\big)\,,
\end{equation}
where $M_\phi$ is a monotonic map (increasing when ${\rm Risk}_S[H]>0$ is required and decreasing for ${\rm Risk}_S[H]<0$). Once again (i.e. in broad agreement with the optimal transfer~(\ref{Eq:c-proj_Mx})) we see the optimal hedging product as a monotonic image of the risk scenario product $S$.

\subsection{Hedge-investment duality and risk recycling}

It turns out that hedging and investment products are intimately related. An optimal hedge can be viewed as an optimal investment and vice versa. In this section we clarify this observation. As a bonus, we gain geometric characterization of rational investments.

Consider a rational investor with utility function $U$ and belief $b$. Let $F$ be the optimal product for such an investor, i.e.
\begin{equation}\label{Eq:maxExpectedUtility}
F=\arg \max_{{\cal F}}\int b(x)U\big({\cal F}(x)\big)\,dx\,,\ \ \ {\rm Price}[{\cal F}]=1\,.
\end{equation}
Let us now compare $F$ with the pure hedge product~(\ref{Eq:OptimalPureHedge_GeneralForm}) defined, as discussed above, by the optimization
\begin{equation}\label{Eq:OptimalPureHedge_Detailed}
H = \Big(\arg \min_\beta D_\phi(\beta\,||\,m)\Big)/m\,,\ \ \ {\rm Risk}_{S}[\beta]=r\,.
\end{equation} 
To line up~(\ref{Eq:OptimalPureHedge_Detailed}) with~(\ref{Eq:maxExpectedUtility}) we recall the definition of the likelihood product~(\ref{Eq:b=fm}) and set 
\begin{equation}
\phi(x)=-U(x)+U(1)\,;\ \ \ S\propto \exp(-1/f)\,,\ \ r={\rm Risk}_S[F]\,.
\end{equation}  
Under these settings the pure hedging product~(\ref{Eq:OptimalPureHedge_Detailed}) coincides with the given rational investment~(\ref{Eq:maxExpectedUtility}), i.e. $H=F$ (see Appendix~\ref{Sec:HedgeInvestmentDuality}).

This observation provides geometric characterization of rational investments. The phenomenon of risk aversion (which is normally captured by $U$) is described geometrically as minimising the divergence $D_\phi(\beta\,||\,m)$ which pushes the traded view $\beta$ towards the market $m$. The investor's real view acts as a constraint. The net behaviour agrees with the expected utility maximization~(\ref{Eq:maxExpectedUtility}).

Above we took a given optimized investment and showed how it can be viewed as a hedge. Let us now explore the opposite direction, i.e. let us start from a hedging problem and see if it can be solved by trading investment products. In doing so we touch upon the very important practical topic of risk recycling.

Imagine being long an asset with some payoff function $A(x)$. We compute the exposure ${\rm Risk}_S[A]$ of the asset with respect to a certain product $S$, and want to learn how to eliminate this exposure through the selling of investment products. We choose $D_\phi$ and design a pure hedge $H$ with the same exposure ${\rm Risk}_S[H]={\rm Risk}_S[A]$. Selling $H$ would obviously eliminate the unwanted risk; we just need to identify appropriate investors. 

Although mathematics cannot guarantee physical presence of such investors on the actual market, we can check the theoretical possibility. To do so we would need to find a possible view $b$ together with the risk aversion profile $R$ which happen to lead to $H$ as an optimal investment product (via the structuring equations~(\ref{Eq:b=fm}) and (\ref{Eq:PayoffElasticity})). 

In the case of positive exposure, ${\rm Risk}_S[A]>0$, we look for investors into the risk scenario $b=\beta_S=S\,m$. In the notation of Eq.~(\ref{Eq:b=fm}) this is equivalent to setting the likelihood product $f=S$. The risk aversion profile is easy to check by computing the payoff elasticity
\begin{equation}\label{Eq:HedgeElasticity}
\frac{d\,\ln H}{d\,\ln f}=\frac{\mu}{H\phi''(H)}\,,
\end{equation}
where $\mu$ is a positive constant (see Appendix~\ref{Sec:HedgeInvestmentDuality}). Comparison with Eq.~(\ref{Eq:PayoffElasticity}) implies a positive risk aversion ($R=H\phi''(H)/\mu>0$). This shows that $H$, which was originally designed as a pure hedging product, may be bought by rational investors. Such investors can be searched for, since we know their example belief $b$ and risk aversion $R$.

In the case of negative exposure, ${\rm Risk}_S[A]<0$, we look for investors with a view that is \lq\lq opposite'' to the scenario $\beta_S$. To this end we introduce a reciprocal of $S$
\begin{equation}\label{Eq:ReciprocalS}
\bar{S}\stackrel{\rm def}{=} \frac{1/S}{{\rm Price}[1/S]}\,.
\end{equation} 
Directly from the definition~(\ref{Eq:SpecificRisk}) we see that ${\rm Risk}_{\bar{S}}[A]=-{\rm Risk}_S[A]$ for any $A$. Setting $b=\beta_{\bar{S}}=\bar{S}\,m$ (i.e. $f=\bar{S}$) and repeating the above arguments, we see that, even in the case of negative exposure, one can theoretically find rational investors who would be interested in buying $H$ from us.

In summary, it is always possible to package an unwanted risk from our inventory in a rational investment product. The marketing materials for such products could contain the implied investors views and risk aversion profiles, thus achieving maximum transparency. Whether such investors would be found depends, of course, on the market.

\subsection{Partially hedged investment products}
\label{Sec:PartiallyHedgedProducts}

Trading logistics is an important factor which can influence product design. A good example is the so-called swap format, where the counterparties enter a contract that is balanced, requiring no capital exchange at inception. Originally inspired by the basic delta-one products such as swaps and futures, this format gained considerable popularity, even among exotics. A very considerable portion of autocallables, for example, is currently traded in the swap format.

In this section we point to the possibility of crafting products with zero initial delta. More generally, a number of first-order risks can be set to desired initial values. Like the swap format, this can ease trading logistics at inception.

Mathematically, such a possibility hinges on the fact that the generalized notion of risk~(\ref{Eq:SpecificRisk}) leads to linear constraints which are easy to incorporate. For example, let us add a risk constraint to the optimization~(\ref{Eq:maxExpectedUtility}), so that
\begin{equation}\label{Eq:RiskConstrainedMaxExpectedU}
F=\arg \max_{{\cal F}}\int b(x)U\big({\cal F}(x)\big)\,dx\,,\ \ \ {\rm Price}[{\cal F}]=1\,,\ \ \ {\rm and}\ \ \ {\rm Risk}_S[{\cal F}]=r\,.
\end{equation}
For the solution of the above optimization we immediately derive (see Appendix~\ref{Sec:PartiallyHedgedInvestments})
\begin{equation}\label{Eq:RiskConstrainedPayoffElasticity}
\frac{d\,\ln F}{d\,\ln f}=\frac{1}{R^*}\,,\ \ \ R^*=R\cdot\Big(1-\frac{d\,\ln (1+\alpha_r\ln S)}{d\,\ln f}\Big)^{-1}\,,
\end{equation}
where $R$ is the familiar Arrow-Pratt relative risk aversion and $\alpha_r$ is chosen to satisfy the risk constraint in~(\ref{Eq:RiskConstrainedMaxExpectedU}). In other words, whenever the risk-constrained optimization~(\ref{Eq:RiskConstrainedMaxExpectedU}) is feasible, the solution satisfies the payoff elasticity equation with a modified risk aversion profile~(\ref{Eq:RiskConstrainedPayoffElasticity}).

\section{Summary and outlook}

Multiple rationality (grounded in neuroscience) explains why we have many different products (each with its own rationale). This observation allows us to use rational optimizations at the level of individual products (information derivatives).

Thinking about financial products we discovered an underlying geometric structure that supports both hedging and investments. At the top (conceptual) level, this gives us a geometric description of rational behaviour (a geometric analogue of the expected utility theory). Concepts such as risk aversion and hedging become clarified in this new light. Risk aversion pushes traded views towards the market-implied consensus. Hedging is similar yet with a clear difference -- it pushes towards a zero-risk surface (which contains the market view as a single point).

At a more detailed level, we understood how to interpret sensitivities as investment spreads, learned how to capture risks using geometric configurations, investigated the general structure of optimal hedging products (both pure hedging and partially hedged products), showed how to quell risks by purchasing insurance assets and how to recycle risks by packaging them into rational investments for sale to suitable clients.

Each of the above results can lead to important developments. Even just the translation of risks into geometric structures may already open interesting perspectives. Indeed, think of the existing technologies behind automatic image recognition or self-driving cars. Casting risks in geometric terms opens the possibility of using established AI techniques to recognize risk configurations and analyse their evolution over time.

Purely mathematical developments are also easy to foresee. It would be natural, for instance, to explore an even more general definition of specific risk by taking Eq.~(\ref{Eq:RiskPythagoras}) and replacing $D$ with a more general Bregman divergence~\cite{Bregman_1967}. 

While we want to encourage a wide range of explorations, it would be prudent to draw the reader's attention to an important scientific issue. Several times in the paper we encountered the measure of risk aversion. We treated it as a mathematical quantity, requiring only that it should be positive (to exclude gambling behaviours). In reality, of course, there must be additional (much more subtle) limits on the types of risk aversion that can be found in nature. Deeper integration with econometrics and neuroscience is required.

\begin{center}
    \rule[1ex]{.25\textwidth}{.5pt}
\end{center}
\vspace*{1cm}

\renewcommand{\thesubsection}{A.\arabic{subsection}}
\section*{Appendix}

\subsection{Multiple rationality}
\label{Sec:MultipleRationality}
In this note we briefly revisit the old question of rationality using an interdisciplinary vantage point of neuroeconomics~\cite{GlimcherFehr_2014}. We argue that products (financial or otherwise) decompose people's overall behaviour into smaller more primitive goals. Upon such a decomposition, the usual diagnosis of irrationality loses some of its meaning allowing room for rational product designs.   

Behavioural economists demonstrated that the overall behaviour of a person frequently deviates from that of an imaginary rational agent. However, as the catalogue of irrational deviations grew so did the realization that the observed deviations do have reasons. The use of heuristics, for example, saves valuable time and energy. Given that the brain can consume up to 20\% of all our power (50\% in children) \cite{Sokoloff_1996} it may be misleading to brand resource-saving heuristics as irrational.

Something more complex and interesting is going on. We see our brains constantly and simultaneously engaging in multiple optimizations just to keep us alive. Take, for instance, functions such as hearing, speech, vision, controls of balance, temperature, breathing, appetite or blood pressure. Note that a great variety of functions are clearly distinct in terms of their core specialized goals (this is in fact what allows us to classify them as separate functions).

The multiplicity of goals inside a single brain can be seen on a physical level. Indeed, neurologists routinely speak about physically distinct brain areas serving this or that particular function (e.g. Brodmann areas~\cite{Brodmann_1909}). Accurate localization of different brain functions is a key goal in modern neuroanatomy with practical applications in surgery.

We use the term {\it multiple rationality} to describe the phenomenon of multiple goals coexisting within a single brain. This does not imply any consistency or harmony between the goals. In fact we know that the individual goals can often be contradictory. Contrast, for instance, the classic pair of behaviours: fight vs flight. Such behaviours rely on very different skills and have completely different criteria of success, and yet every person has some capacity in both (physically implemented within the relevant brain structures).

Multiple rationality is a general neurological phenomenon which goes beyond our species. It was found, for instance, that the brain of a desert ant looking for a way home simultaneously computes different navigation strategies~\cite{WystrachEtAl_2013}. Remarkably, even in such a small brain, where every neuron is precious, millions of years of evolution settled on multiple rationality.

Probabilistic reasoning is at the heart of this paper, and it is a particular aspect of rationality which has been challenged by behavioural studies~\cite{TverskyKahneman_1983}. Multiple rationality suggests that, indeed, it might be very difficult to detect accurate probabilistic reasoning in a subject unless we have sufficient (e.g. surgical) control of the experimental conditions. In Ref.~\cite{YangShadlen_2007} the authors report an experiment in which rhesus monkeys had to make decisions under uncertainty while the signals from the relevant neurons were recorded via surgically implanted electrodes. The recorded signals were shown to be proportional to the log-likelihood ratios. While not many economists would think of monkeys as rational agents, suitably isolated areas of the monkeys' brains were observed performing as expert statisticians.

In summary, big-picture views on human behaviour mask and distort the underlying decision-making mechanics. In reality, the brain is an incredibly complex network of multiple decision-making centres some of which could be collaborating, competing, or preparing for action at a later time. Sometimes an individual goal can be isolated and addressed with a suitably designed product (e.g. a pair of glasses improving clarity of vision). This opens up the possibility of rational product designs.

\subsection{$b=fm$, and the Bayesian inference hierarchy}
\label{Sec:InferenceProduct}

Our more mathematical readers often note the connection between the likelihood product (\ref{Eq:b=fm}) and the measure change between market-implied and investor-believed distributions. This observation is interesting for two reasons. Firstly, the concept of measure change, which is ubiquitous in financial mathematics, becomes connected to product design. For a discussion of this aspect see Ref.~\cite{Soklakov_2013a}. Secondly, significant mathematical generality of the information derivatives framework is ensured. Indeed, it is important that the framework can accommodate a great variety of ways in which the believed distribution can be obtained (including machine learning, etc.). The conditions for existence of a likelihood product are essentially the same as that for the equivalent Radon-Nikodym derivative (i.e. extremely general).

While mathematical generality ensures compatibility with most modern and even future inference algorithms, it could be helpful to develop some intuition using classical examples. Such intuition also helps understanding information derivatives in relation to other inference-based financial optimizations such as the Black-Litterman framework~\cite{BlackLitterman_1990}.

With this pedagogical end in mind, we allow ourselves a brief review of the Bayesian inference hierarchy. The bulk of this section is very well known in the inference literature even if the relevant financial applications are more specialized. A curious reader may note that Bayesian inference can itself be viewed as an extension of logic~\cite{Cox_1946, Jaynes_2003}, so the below examples have indeed a very solid theoretical foundation. 

\subsubsection{Bayesian inference}
We use the term {\it inference} to mean a process of updating our prior knowledge to a posterior on the account of learning new information. Bayesian probability updating is perhaps the simplest such example. Bayesian updating of the prior $P(x)$ to a posterior $P^*(x)$ on the account of observing data $D$ consists of two parts
\begin{eqnarray}
&{\rm Bayes'\ rule:}\ \ \ P^*(x) = &\!\!\!P(x|D)\,, \label{Eq:Bayes'Rule}\\
&{\rm and\ Bayes'\ theorem:}\ \ &\!\!\!P(x|D) = \frac{P(D|x)}{P(D)}P(x)\,. \label{Eq:Bayes'Theorem}
\end{eqnarray} 
The resulting inference takes the product form of Eq.~(\ref{Eq:b=fm})
\begin{equation}
P^*(x)= L_B(x)P(x)\,,\ \ {\rm where\ Bayesian\ likelihood}\ \ L_B(x)= \frac{P(D|x)}{P(D)}\,.
\end{equation} 

\subsubsection{Jeffrey's inference}
Bayes' theorem is a well known mathematical fact. It follows from the internal consistency of probability assignments~\cite{Cox_1946}. Jeffrey recognized Bayes' rule as a separate logical step within inference and argued that a more general principle was needed~\cite{Jeffrey_1965}. 

Consider for instance an important practical situation when all new information comes from an imperfect measurement apparatus. The apparatus is unable to return the true value $y$ and measuring $y_0$ means that the true value is somewhere near $y_0$ as described by the distribution $P^*_{y_0}(y)$. The posterior distribution in this case is given by
\begin{eqnarray}
&{\rm Jeffrey's\ rule:}\ \ \ \ P^*(x) = \int &\!\!\!\! P(x|y) P^*_{y_0}(y)\,dy \,, \label{Eq:Jeffrey'sRule}\\
&{\rm where\ by\ Bayes'\ theorem:}\ \ &\!\!\!\!P(x|y) = \frac{P(y|x)}{P(y)}P(x)\,. \label{Eq:Jeffrey'sBayes}
\end{eqnarray} 
In the special case of an ideal measurement $P^*_{D}(y)=\delta(y-D)$ this reduces to Bayesian inference~(\ref{Eq:Bayes'Rule}, \ref{Eq:Bayes'Theorem}). We want, however, to cast Jefferey's inference into the product form of Eq.~(\ref{Eq:b=fm}) in full generality. Substituting (\ref{Eq:Jeffrey'sBayes}) into (\ref{Eq:Jeffrey'sRule}) we derive
\begin{equation}
P^*(x)= L_J(x)P(x)\,,\ \ {\rm where\ Jeffrey's\ likelihood}\ \ L_J(x)= \int P^*_{y_0}(y)\frac{P(y|x)}{P(y)}\,dy\,.
\end{equation} 

The Black-Litterman model for asset allocation~\cite{BlackLitterman_1990} is an excellent, and relevant for our purposes, example of a practical application of the Bayesian approach (Jeffrey's style) in finance~\cite{KolmRitter_2017}. The original Black-Litterman portfolios can be viewed as information derivatives which are subject to constraints and assumptions: the payoffs are restricted to linear combinations of assets; all new information is about linear combinations of returns; risk aversion is modelled by a quadratic utility (mean-variance optimization of returns); and all distributions (priors and likelihoods) are assumed to be Gaussian. Understanding the Bayesian structure of Black-Litterman is a major step in its generalizations~\cite{KolmRitter_2017}.

\subsubsection{Entropic inference methods}
Sometimes new information is discovered in the form of constraints. In such cases, the Jaynes principle of maximum entropy or, more generally, Kullback's principle of minimal divergence are often used (see~Ref.~\cite{ShoreJohnson_1980} and references therein). The relevant rule for updating from the prior $P$ to the posterior $P^*$ reads as a conditional optimization
\begin{equation}\label{Eq:MaximumEntropyRule}
P^*=\arg\min_q \int q(x)\ln\frac{q(x)}{P(x)}\,dx\,,
\end{equation}
where all $\{q\}$ are subject to the constraints which convey new information. 

For instance, one might learn that the expected value of some function $g(x)$ should be equal to $\bar{g}$. Such expectation-constrained optimizations have been known in statistical physics for over a century. The ubiquity of the resulting probability distributions in physics earned them the name \lq\lq canonical''. From Eq.~(\ref{Eq:MaximumEntropyRule}) we immediately derive
\begin{equation}
P^*(x)= L_C(x)P(x)\,,\ \ {\rm where\ the\ canonical\ likelihood}\ \ L_C(x)\propto e^{cg(x)}\,,
\end{equation}
and $c$ is a constant (Lagrange multiplier) which is fixed to satisfy the constraint. The correspondence with Eq.~(\ref{Eq:b=fm}) is evident and the likelihood product is easy to obtain. 

An excellent discussion of the variational formulation~(\ref{Eq:MaximumEntropyRule}) with a few additional examples can be found in Ref.~\cite{Caticha_2012}. For an example of a relevant financial application we may continue the theme of upgrading Black-Litterman~\cite{MeucciEtAl_2014}.
 
For completeness, the maximum entropy (the minimum divergence) inference~(\ref{Eq:MaximumEntropyRule}) includes the Bayesian and Jeffrey's  inference as special cases~\cite{Williams_1980}. A further generalization in which the extremal distribution~(\ref{Eq:MaximumEntropyRule}) is just a leading contributor to inference is considered in Ref.~\cite{Caticha_2000}. The exploration of the relevant likelihood products we leave to the reader as an easy exercise.

\subsection{Specific risk as a spread in expected returns}
\label{Sec:SpecificRisk}
From Eqs.~(\ref{Eq:Sensitivity}) and (\ref{Eq:ScoreProduct}) we derive
\begin{equation}\label{Eq:SpecificSensitivity_score}
\frac{\partial_\sigma{\rm Price}[\Pi]}{{\rm Price}[\Pi]}\, =\, \int \beta_\Pi(x)\, {\rm Score}(x)\,dx\, \equiv\, E_{\beta_\Pi}[{\rm Score}]\,,
\end{equation}
where
\begin{equation}
{\rm Score} = \ln F_0 +\ln {\rm Price}[e^{\rm Score}]\,.
\end{equation}

Since the $m_\sigma$-expectation of ${\rm Score}$ is zero we proceed
\begin{eqnarray}
{\rm Score} & = &\ln F_0 +\ln {\rm Price}[e^{\rm Score}]-E_{m_\sigma}[{\rm Score}]\cr
&&\cr
&=& \ln F_0 +\ln {\rm Price}[e^{\rm Score}]-E_{m_\sigma}[\ln e^{\rm Score}] \cr
&&\cr
&=& \ln F_0 -E_{m_\sigma}\Big[\ln e^{\rm Score} - \ln {\rm Price}[e^{\rm Score}]\Big] \cr
&&\cr
&=& \ln F_0 -E_{m_\sigma}[\ln F_0]\,.
\end{eqnarray}
Substituting this into Eq.~(\ref{Eq:SpecificSensitivity_score}) we obtain Eq~(\ref{Eq:SpecificSensitivity_spread}).\\

\subsection{Specific risk in terms of relative entropy}
\label{Sec:SpecificRisk_RelativeEntropy}

Written out explicitly Eq.~(\ref{Eq:SpecificRisk}) reads
\begin{eqnarray}
\frac{{\rm Risk}_S[\Pi]}{{\rm Price}[\Pi]} &\stackrel{\rm def}{=}& \int \beta_{\Pi}(x) \ln S(x)\,dx - \int m(x) \ln S(x)\,dx\cr
&& \cr
&=& \int \beta_{\Pi}(x) \ln \frac{\beta_S(x)}{m(x)}\,dx - \int m(x) \ln \frac{\beta_S(x)}{m(x)}\,dx\,,\ \ {\rm where}\ \  \beta_S=Sm.\ \ \ \ \ \ 
\end{eqnarray}

We proceed by direct calculation
\begin{eqnarray}
\frac{{\rm Risk}_S[\Pi]}{{\rm Price}[\Pi]} &=& \int \beta_{\Pi}(x) \ln \frac{\beta_\Pi(x)}{m(x)} \frac{\beta_S(x)}{\beta_\Pi(x)}\,dx + \int m(x) \ln \frac{m(x)}{\beta_S(x)}\,dx\\
&& \cr
&=& \int \beta_{\Pi}(x) \ln \frac{\beta_\Pi(x)}{m(x)}\,dx + \int m(x) \ln \frac{m(x)}{\beta_S(x)}\,dx\,-\int \beta_{\Pi}(x) \ln \frac{\beta_\Pi(x)}{\beta_S(x)}\,dx\,, \nonumber
\end{eqnarray}
which proves Eq.~(\ref{Eq:RiskPythagoras}).\\

\subsection{Financial meaning of geodesics}
\label{Sec:GeodesicsFinMeaning}

\subsubsection{M-geodesics}
Imagine derisking the portfolio $\Pi$ by converting $(1-t)$-fraction of it into cash. Assuming this can be done with negligible loss, we compute the payoff function of the resulting new portfolio 
\begin{equation}
\Pi_t(x) = (1-t){\rm Price}[\Pi]+t\Pi(x)\,.
\end{equation}
Examining the corresponding growth-optimal view we derive
\begin{eqnarray}
\beta_{\Pi_t}(x) &=& \frac{\Pi_t(x)}{{\rm Price}[\Pi_t]}\,m(x) \cr
&&\cr
&=& (1-t)m(x) +t\beta_\Pi(x)\cr
&&\cr
&=& p_{\rm mix}(x,t)\,.
\end{eqnarray}
In other words, moving along the m-geodesic $p_{\rm mix}(x,t)$ from $\beta_\Pi$ to $m$ describes proportional liquidation of the portfolio.

\subsubsection{E-geodesics}

From the definition~(\ref{Eq:e-geodesic}), the e-geodesic $p_{\rm exp}(x,t)$ can be written as the family of different geometric averages between $m$ and $\beta_S$   
\begin{equation}
p_{\rm exp}(x,t) \propto m^{1-t}(x)\beta_S^t(x)\,.
\end{equation}
Recalling the relationship between the risk product $S$ and the risk scenario $\beta_S=Sm$
\begin{equation}\label{Eq:e-geodesic-S}
p_{\rm exp}(x,t)=\frac{m(x)\,S^t(x)}{\int m(y)\,S^t(y)\,dy }\,.
\end{equation}
Moving along the e-geodesic is equivalent to replacing the original risk product $S$ with 
\begin{equation}
S_t(x)=p_{\rm exp}(x,t)/m(x)\propto S^t(x).
\end{equation}
In order to compare $S$ with $S_t$ we recall the payoff elasticity equation~(\ref{Eq:PayoffElasticity}) in the relative form~\cite{Soklakov_2013b}
\begin{equation}\label{Eq:RelativePayoffElasticity}
\frac{d\ln F_1}{d\ln F_2}=\frac{R_2}{R_1}\,,
\end{equation} 
where $F_1$ and $F_2$ is a pair of investment products corresponding to the pair of risk aversion profiles $R_1$ and $R_2$. Setting $F_1=S_t$ and $F_2=S$ we compute 
\begin{equation}
\frac{R_2}{R_1}=\frac{d\ln S_t}{d\ln S}=t\,.
\end{equation} 
i.e. moving along the e-geodesic is equivalent to scaling risk aversion (on the risk product).

\subsection{Exposure along the risk geodesic}
\label{Sec:ExposureRiskGeodesic}

Note that the rhs of~(\ref{Eq:RiskPythagoras}) contains only distributions. This fact is interesting because it leads to geometric pictures that are invariant to portfolio size. It is convenient to overload our notation of risk so we can use it at the level of distributions. We define
\begin{equation}
{\rm Risk}_S[\beta_\Pi]\stackrel{\rm def}{=}\frac{{\rm Risk}_S[\Pi]}{{\rm Price}[\Pi]}\,.
\end{equation}
In this paper we are not interested in exploring special pathological examples and so we can assume that the rhs of Eq.~(\ref{Eq:e-geodesic-S}) is always well defined including the values of $t\notin[0,1]$. Let $\beta_t(x)$ be the extrapolated version of the e-geodesic $p_{\rm exp}(x,t)$ obtained by removing any restriction on the values of $t$ in Eq.~(\ref{Eq:e-geodesic-S}). We compute
\begin{eqnarray}
{\rm Risk}_S[\beta_t]&=&\int\Big(\beta_t(x)-m(x)\Big)\Big(\ln\beta_S(x)-\ln m(x)\Big)\,dx\cr
&&\cr
&=&\int \Big(\frac{m(x)\,S^t(x)}{\int m(y)\,S^t(y)\,dy }-m(x)\Big)\ln S(x)\,dx\,.
\end{eqnarray}

Differentiating with respect to the parameter $t$ and re-combining the expressions
\begin{eqnarray}
\frac{d}{dt}{\rm Risk}_S[\beta_t] &=& \frac{\int m(x)\,S^t(x)\ln^2 S(x)\,dx}{\int m(y)\,S^t(y)\,dy }-\frac{\big(\int m(x)\,S^t(x)\ln S(x)\,dx\,\big)^2}{\big(\int m(y)\,S^t(y)\,dy\, \big)^2}\cr
&&\cr
&=&\int\beta_t(x)\,\ln^2S(x)\,dx - \Big(\int \beta_t(x)\ln S(x)\,dx\Big)^2\cr
&&\cr
&=&{\rm Var}_{\beta_t}[\ln S]\,,
\end{eqnarray}
where the last equality is really just the standard definition of variance. Since the variance is positive we see the exposure along the risk geodesic changes monotonically (increasing in the direction from $m$ to $\beta_S$). Integrating this back gives us Eq.~(\ref{Eq:RiskAsIntegratedVar}).

\subsection{Iso-risk foliation}
\label{Sec:IsoRiskFoliation}
In Appendix~\ref{Sec:ExposureRiskGeodesic} we found that the exposure to $S$ along the e-geodesic $|m,\beta_S\rangle$ changes monotonically. This means that the e-geodesic intersects all the iso-risk hyper-surfaces defined by fixing different values of ${\rm Risk}_S$. Here we show that all such intersections happen at the right angles.  

Select a value of risk $r$ and consider the iso-risk surface $\Omega_r$ formed by points $\beta_\Pi$ such that ${\rm Risk}_S[\beta_\Pi]=r$. Let $\beta_{t_r}$ be the point of intersection between the e-geodesic $|m,\beta_S\rangle$ and $\Omega_r$. By construction for any arbitrary $\beta_\Pi \in \Omega_r$ we have
\begin{equation}
{\rm Risk}_S[\beta_\Pi]={\rm Risk}_S[\beta_{t_r}]\,,
\end{equation}
or equivalently
\begin{equation}\label{Eq:RiskEquality}
\int\Big(\beta_\Pi(x)-m(x)\Big)\ln\frac{\beta_S(x)}{m(x)}\,dx =\int\Big(\beta_{t_r}(x)-m(x)\Big)\ln\frac{\beta_S(x)}{m(x)}\,dx\,.
\end{equation}
We want to inspect the intersection angle $\angle\beta_\Pi\beta_{t_r}\beta_S$. To this end we compute
\begin{eqnarray}
\langle\beta_{t_r}, \beta_\Pi|\beta_{t_r},\beta_S\rangle
&=&\int\Big(\beta_\Pi(x)-\beta_{t_r}(x)\Big)\ln\frac{\beta_S(x)}{\beta_{t_r}(x)}\,dx\cr
&&\cr
&=& \int\Big(\beta_\Pi(x)-m(x)\Big)\ln\frac{\beta_S(x)}{\beta_{t_r}(x)}\,dx + \int\Big(m(x)-\beta_{t_r}(x)\Big)\ln\frac{\beta_S(x)}{\beta_{t_r}(x)}\,dx\cr
&&\cr
&=& \int\Big(\beta_\Pi(x)-m(x)\Big)\Big(\ln\frac{\beta_S(x)}{m(x)}- \ln\frac{\beta_{t_r}(x)}{m(x)}\Big)\,dx\cr
&& + \int\Big(m(x)-\beta_{t_r}(x)\Big)\Big(\ln\frac{\beta_S(x)}{m(x)}-\ln\frac{\beta_{t_r}(x)}{m(x)}\Big)\,dx
\end{eqnarray}
Noticing cancellations due to Eq.~(\ref{Eq:RiskEquality}) we proceed
\begin{eqnarray}
\langle\beta_{t_r}, \beta_\Pi|\beta_{t_r},\beta_S\rangle
&=&\int\Big(\beta_{t_r}(x)-m(x)\Big)\ln\frac{\beta_{t_r}(x)}{m(x)}\,dx-\int\Big(\beta_\Pi(x)-m(x)\Big)\ln\frac{\beta_{t_r}(x)}{m(x)}\,dx\cr
&&\cr
&=& \int\Big(\beta_\Pi(x)-\beta_{t_r}(x)\Big)\ln\frac{m(x)}{\beta_{t_r}(x)}\,dx\cr
&&\cr
&=&\langle\beta_{t_r}, \beta_\Pi|\beta_{t_r}, m\rangle\,.
\end{eqnarray}
This means that the angles $\angle\beta_\Pi\beta_{t_r}\beta_S$ and $\angle\beta_\Pi\beta_{t_r}m$ are of the same type (both acute, both obtuse, or both right). Since $m$, $\beta_{t_r}$ and $\beta_S$ lie on the same geodesic, the two angles are complimentary and therefore can only be both right. This completes the proof that the risk geodesic is orthogonal to the corresponding iso-risk hyper-surfaces.

The collection of all iso-risk surfaces constitutes an m-foliation (meaning that every iso-risk surface is m-flat and together they span the entire manifold of probability distributions). Orthogonal to that is an e-foliation. In the above example the risk geodesic $|m,\beta_S\rangle$ is an element of an e-foliation. Such dual foliations are useful in defining convenient coordinate systems in which some important directions look straight (as in Figs.~1 and~2). For a detailed mathematical treatment of dual foliations and mixed coordinates see Sec.~6.8 of~\cite{Amari_2016}.

\subsection{Optimizations}
\subsubsection{C-projection}
\label{Sec:C-projection}

The Lagrangian for finding the c-projection $\beta_\Pi\to\beta_{\Pi'}$ onto the zero-risk manifold with minimal cost~(\ref{Eq:Cost}) reads
\begin{equation}
{\cal L}_{\Pi'}= \int C\big( x,\beta_{\Pi'}(x)-\beta_\Pi(x) \big)\, dx -\lambda_c\cdot\big({\rm Price}[\Pi']-1\big) - \mu_c\cdot{\rm Risk}_S[\Pi']\,,
\end{equation}
where $\lambda_c$ and $\mu_c$ are the Lagrange multipliers ensuring the correct normalization and the zero risk for $\beta_{\Pi'}$. Rearranging the terms
\begin{equation}
{\cal L}_{\Pi'}= \int \Big[ C\big( x,\beta_{\Pi'}(x)-\beta_\Pi(x) \big)-\lambda_c\beta_{\Pi'}(x) - \mu_c\beta_{\Pi'}(x)\ln S(x)\Big]\,dx+\dots\,,
\end{equation}
where the hidden terms do not depend on $\Pi'$. Variation with respect to $\beta_{\Pi'}$ gives the Euler-Lagrange equation
\begin{equation}
C'_2\big(x,\beta_{\Pi'}(x)-\beta_\Pi(x)\big)=\lambda_c+\mu_c\ln S(x)\,,
\end{equation} 
where $C'_2$ denotes the derivative of $C$ with respect to the second argument. Since $C$ is convex with respect to its second argument, the above equation can be inverted locally (i.e. at every $x$) to expose the difference
\begin{equation}
\beta_{\Pi'}(x)-\beta_\Pi(x)=C'^{-1}_2\big(x,\lambda_c+\mu_c\ln S(x)\big)\,,
\end{equation} 
where $C'^{-1}_2$ is monotonic with respect to its second argument. Dividing both sides by $m(x)$ we arrive at Eq.~(\ref{Eq:c-proj_Mx}).

\subsubsection{Pure investment products}
\label{Sec:Pure_investment_products}
By a {\it pure investment product} we mean a solution $F$ to the problem of maximum expected utility~(\ref{Eq:maxExpectedUtility}). Within the framework of information derivatives such products are structured by solving the pair of equations~(\ref{Eq:b=fm}) and (\ref{Eq:PayoffElasticity}). This brings into play the key notions of likelihood and risk aversion. Here we recall an alternative form of writing the same result (more standard in mathematics). This is in preparation for understanding the hedge-investment duality where we use both forms (Appendix~\ref{Sec:HedgeInvestmentDuality}).

The Lagrangian for finding the optimal investment product $F$~(\ref{Eq:maxExpectedUtility}) reads
\begin{equation}
{\cal L}_F=\int b(x)\,U\big(F(x)\big)\,dx - \lambda_0\,\Big(\int m(x) F(x)\,dx-1\Big)\,,
\end{equation}
where $\lambda_0$ is the Lagrange multiplier. The corresponding Euler-Lagrange equation
\begin{equation}
b(x)\,U'\big(F(x)\big)-\lambda_0 m(x)=0
\end{equation}
immediately gives us the solution
\begin{equation}\label{Eq:F_ClassicSolution}
F(x)=U'^{-1}\big(\lambda_0/f(x)\big)\,,\ \ \ {\rm where}\ \ \ f(x)=b(x)/m(x)\,,
\end{equation}
$U'^{-1}$ is the inverse function of $U'$.

\subsubsection{Pure hedging products -- the general structure}
\label{Sec:Pure_hedging_products_gen_structure}
As explained in the main text, a {\it pure hedging product} is defined as a solution $H$ of the divergence minimization problem with the Lagrangian
\begin{equation}\label{Eq:D_phi_Lagrangian}
{\cal L}_H = D_\phi(\beta_H\,||\,m)-\lambda\cdot({\rm Price}[H]-1)-\mu\cdot({\rm Risk}_{S}[H]-r)\,,
\end{equation}
where $\lambda$ and $\mu$ are the Lagrange multipliers for the constraints on the price and the risk of $H$ respectively. The intuition behind this is very simple: a unit of a pure hedging product is designed to express as little view on the market as possible -- just what it needs to have the required risks. More explicitly the Lagrangian reads 
\begin{equation}
{\cal L}_H=\int m(x)\Big[\phi\big(H(x)\big)-\Big(\lambda+\mu\ln S(x)\Big)H(x)\Big]\,dx+\dots\,,
\end{equation}
where the hidden terms do not depend on $H$. The variation with respect to $H$
\begin{equation}
\delta_H{\cal L}_H=\int m(x)\Big[\phi'\big(H(x)\big)-\Big(\lambda+\mu\ln S(x)\Big)\Big]\delta H(x)\,dx\,,
\end{equation}
and the Euler-Lagrange equation
\begin{equation}
\phi'\big(H(x)\big)=\lambda+\mu\ln S(x)\,.
\end{equation}
For a strictly convex function $\phi$, the derivative $\phi'$ is monotonically increasing and therefore invertible. We can therefore compute
\begin{equation}\label{Eq:H*}
H(x)=\phi'^{-1}\big(\lambda+\mu\ln S(x)\big)\,,
\end{equation}
where the inverted function $\phi'^{-1}$ is itself monotonically increasing. The values $\lambda$ and $\mu$ are determined from the constraints (on the price and risk of $H$). 

Equation~(\ref{Eq:H*}) shows that the optimal hedging product is essentially the same as the risk product. More precisely, $H$ is a monotonic function of $S$. Below, in the second half of note~\ref{Sec:HedgeInvestmentDuality}, we show this function to be monotonically increasing if the positive exposure is required and monotonically decreasing if we want ${\rm Risk}_S[H]<0$.

\subsubsection{Hedge-investment duality}
\label{Sec:HedgeInvestmentDuality}

{\bf Investment $\to$ Hedge}\\ % {\bf Geometry of rational investment}
Let $F$ be a pure investment product of the form (\ref{Eq:F_ClassicSolution}). We want to see if we can find the optimal pure hedge of the form~(\ref{Eq:H*}) which coincides with $F$.

If we choose $\phi(x)=-U(x)+U(1)$, Eq.~(\ref{Eq:H*}) becomes
\begin{equation} \label{Eq:HU'-1}
H(x)=U'^{-1}\big(-\lambda-\mu\ln S(x)\big)\,.
\end{equation}
Comparison to (\ref{Eq:F_ClassicSolution}) suggests a possible choice of $S=c\cdot\exp(-1/f)$, where $c$ is the notional constant $c=1/{\rm Price}[\exp(-1/f)]$. Substituting this into~(\ref{Eq:HU'-1}) we compute
\begin{equation}
H(x)=U'^{-1}\big(-\lambda-\mu\ln c + \mu/f(x)\big)\,.
\end{equation}
This would coincide with Eq.~(\ref{Eq:F_ClassicSolution}) as long as $\mu=\lambda_0$ and $\lambda=-\lambda_0\ln c$. We just need to understand the circumstances in which $\mu$ and $\lambda$ take these exact values. This happens if we equate the risks of $H$ and $F$, i.e. if we require ${\rm Risk}_S[H]={\rm Risk}_S[F]$. Indeed, $\lambda$ and $\mu$ are the Lagrange multipliers which ensure that $H$ is of unit price and has the required amount of risk with respect to $S$. Both constraints are satisfied by $H=F$ vindicating the above values of $\mu$ and $\lambda$.

In summary, for any fixed rational investment $F$ given by the expected utility maximization~(\ref{Eq:maxExpectedUtility}) one can find a pure hedging product $H$ of the form (\ref{Eq:OptimalPureHedge_Detailed}) such that $H=F$.\\

{\bf Hedge $\to$ Investment}\\
The argument in the main part of the paper requires us to compute the derivative $d\ln H/d\ln f$. Before we do that let us go back to Eq.~(\ref{Eq:H*}) and examine it in a bit more detail. In particular, we need to investigate the sign of $\mu$ in Eq.~(\ref{Eq:H*}). This sign determines whether $H$ is an increasing or a decreasing function of $S$.

Let us prove that $\mu$ in Eq.~(\ref{Eq:H*}) has the same sign as the required risk~(\ref{Eq:RiskOfH}). To this end, let us consider a less constrained version of the minimization~(\ref{Eq:D_phi_Lagrangian}) by forgetting about the risk~(\ref{Eq:RiskOfH}). It is easy to see that the solution to this risk-unconstrained version is $H=1$ (which is the same as $\beta_H=m$). Because ${\rm Risk}_S[1]=0$ we conclude that the constraint ${\rm Risk}_S[H]=0$ in the context of our optimization is in fact equivalent to having no constraints on the risk. 

The minimization~(\ref{Eq:D_phi_Lagrangian}) is a strictly convex optimization with linear constraints. Changing the value of $r$ to a nonzero value in~(\ref{Eq:RiskOfH}) moves the constraint hyperplane away from the unique minimum point. By strict convexity this results in a monotonic increase of the optimal objective value (for as long as the constraints remain feasible). In other words, the optimal objective ${\cal L}^*_H$ is an increasing function of $r$ for $r>0$ and a decreasing function of $r$ for $r<0$. 

It now remains to recall the interpretation of the optimal Lagrange multipliers as the derivatives of the objective with respect to the relevant constraints: $\mu=d{\cal L}^*_H/dr$. It follows that $\mu$ in the solution~(\ref{Eq:H*}) is positive for $r>0$ and negative for $r<0$.

The above arguments provide mathematical rigour for an otherwise completely intuitive result that a pure hedge $H$ with a positive (negative) exposure to $S$ is monotonically increasing (decreasing) with $S$. 

Equation~(\ref{Eq:HedgeElasticity}) is derived by differentiating Eq.~(\ref{Eq:H*}) while remembering the sign of $\mu$. 

\subsubsection{Partially hedged investments}
\label{Sec:PartiallyHedgedInvestments}

The Lagrangian for the optimization~(\ref{Eq:RiskConstrainedMaxExpectedU}) reads
\begin{equation}
{\cal L}=\int b(x)U\big(F(x)\big)\,dx-\nu\cdot({\rm Price}[F]-1)-\rho\cdot({\rm Risk}_S[F]-r)\,,
\end{equation}
where $\nu$ and $\rho$ are the Lagrange multipliers. Writing $U(F)=u(\ln F)$ and hiding the terms which do not depend on $F$
\begin{equation}
{\cal L}=\int b(x)u\big(\ln F(x)\big)\,dx - \int m(x)\big(\nu+\rho\ln S(x)\big) F(x)\,dx+\dots\,.
\end{equation}
The variation with respect to $F$
\begin{equation}
\delta_F{\cal L}=\int \Big[ b(x)\frac{u'\big(\ln F(x)\big)}{F(x)} - m(x)\big(\nu+\rho\ln S(x)\big)\Big]\delta F(x)\,dx
\end{equation}
and the Euler-Lagrange equation
\begin{equation}
F=\frac{u'(\ln F)}{\nu+\rho\ln S} f\,,
\end{equation}
where $f=b/m$. Taking the logarithm on both sides and differentiating
\begin{equation}
d\ln F=\frac{u''(\ln F)}{u'(\ln F)}d\ln F  +d\ln f - d\ln(1+\alpha_r\ln S)\,,
\end{equation}
where $\alpha_r=\rho/\nu$. Rearranging the terms
\begin{equation}
\Big(1-\frac{u''(\ln F)}{u'(\ln F)}\Big)\frac{d\ln F}{d\ln f}=1-\frac{d\ln(1+\alpha_r\ln S)}{d\ln f}\,.
\end{equation}
Recognizing the Arrow-Pratt definition of relative risk aversion
\begin{equation}
1-\frac{u''(\ln F)}{u'(\ln F)}=-F\frac{U''(F)}{U'(F)}=R\,,
\end{equation}
we arrive at Eq.~(\ref{Eq:RiskConstrainedPayoffElasticity}).

\end{document}